\documentclass[aps,pre,twocolumn,showpacs]{revtex4}
\usepackage{epsfig}
\usepackage{graphicx}
\usepackage{bm}
\usepackage{amssymb}
\usepackage{amsmath}
\usepackage{amsthm}
\begin{document}
\title{Host--parasite models on graphs}
\author{Matti Peltom\"aki, Ville Vuorinen, Mikko Alava}
\affiliation{
Laboratory of Physics, Helsinki University of Technology, P.O.Box
1100, 02015 HUT, Finland}
\author{Martin Rost}
\affiliation{
Bereich Theoretische Biologie, IZMB, Universit\"at Bonn, Kirschallee
1, 53115 Bonn, Germany}
\pacs{74.60.Ge, 05.40.-a, 74.62.Dh}
\date{\today}

\begin{abstract}
The behavior of two interacting populations, ``hosts'' and
``parasites'', is investigated on Cayley trees and scale-free
networks. In the former case analytical and numerical arguments
elucidate a phase diagram for the  Susceptible-Infected-Susceptible model, 
whose most interesting feature is the
absence of a tri-critical point as a function of the two independent
spreading parameters. For scale-free graphs, the parasite population
can be described effectively by its 
dynamics in a host background. This is shown both by considering the
appropriate dynamical equations and by numerical simulations on
Barab\'asi-Albert networks with the major implication that in the
thermodynamic limit the critical parasite spreading parameter
vanishes. Some implications and generalizations are discussed.
\end{abstract}

\maketitle

\section{Introduction}
Population models, or reaction-diffusion systems have
attracted enormous interest both in the statistical physics
community and as abstract versions of real biological 
dynamics. One particular aspect is the presence of phase
transitions and the contact process or directed percolation
in various disguises (see below, \cite{Haye,MD}). 

Host--parasite or predator--prey systems are a natural extension of
single species models. By their classical results Lotka and Volterra
were able to explain the nature of abundance oscillations in
interacting species \cite{L,V}. In regular landscapes or lattices,
with a finite spreading rate of the species, these oscillations
appear as traveling waves, which can be regular or chaotic, depending
on the interplay of time scales in population dynamics and spreading,
though it is not clear if the phenomenon survives in the thermodynamic
limit \cite{BM,BHS,RoA,DA,AD,CHM,HCM}. In nature they have been
observed in different systems, to name two extreme cases, e.g.\ in
vole populations \cite{RK} and for human diseases such as measles
\cite{CHS}. In the case of measles in a population living on a
landscape of nontrivial island structure, power law fluctuations are
found instead \cite{RA}.

Much of these ideas have recently been generalized in the context of
small-world or in particular ``scale-free'' graphs
\cite{ab02,dmbook03,n03,pvbook04}. For the latter, a perfectly valid
example is given by epidemics of viruses in the Internet since it has
as a graph a fat-tailed probability distribution of the number of
nearest neighbors, $P(k)$. Recently, various models have been studied
as the particulars of the structure --- as the so-called degree
distribution gamma in $P(k) \sim k^{-\gamma}$ --- are varied. A
fundamental discovery concerning disease spreading is an absence of
epidemic threshold in the limit of infinite graphs and the finite-size
effective ``critical point'' obeys an unusual scaling as $L$, the
graph size, is varied \cite{BA,PSV,ML}.

This closely relates to the present work where we study the influence
of a network or graph like structure of the underlying landscape on
host--parasite or predator--prey dynamics. The main findings are (i)
the absence of oscillations, (ii) the absence of an infection
threshold in the limit of an infinite scale free graph, and (iii) the
existence of two {\em separate} transitions in the case of Bethe
lattices with finite coordination number $z$ (``empty'' $\to$ ``hosts
only'', ``hosts only'' $\to$ ``hosts plus parasites'', but {\em no}
transition ``empty'' $\to$ ``hosts plus parasites''). The structure of the rest
of the paper is as follows: Section II contains the necessary
definitions, and the two following ones analytical considerations 
and, to compare, numerical simulations of the models. Finally,
Section V finishes the paper with a discussion.

\section{Model formulation}

\subsection{States and rates}
A basic model for epidemiological applications is the {\em contact
  process}, or the so-called {\em SIS} model. Here one considers 
individuals living on the nodes of an underlying graph which are
either infected ($I$) or susceptible ($S$) to an infection. An
infected individual may spread the disease to a susceptible one if both
are in contact, i.e.\ if they live on neighboring nodes of the
graph. Infected individuals recover with a certain rate and in this
simple version immediately become susceptible for a new infection. So
the dynamics of the {\em SIS} model is defined by the rates
\begin{eqnarray}
r_{S \to I} & = & \lambda \; \; \; \; \; \;  \mbox{if any neighbor is
  infected} 
\nonumber \\
r_{I \to S} & = & 1.
\end{eqnarray}

In this work we generalize the {\em SIS} model to a system with hosts
and parasites ({\em HP}). In other words we consider infections of a
second kind only able to spread onto sites with infections of first
kind. So each node in the graph can be in three possible states: Empty
($e$), populated by a healthy host ($h$), or by a host with parasite
($p$). Between three states there are six possible transitions so the
dynamics are defined by the following rates
\begin{eqnarray}
r_{e \to h} & = & \lambda \; \; \; \; \; \;  \mbox{if any neighbor
  has a host ($h$)} 
\nonumber \\
r_{h \to p} & = & \alpha \; \mu \; \; \; \; \; \;  \mbox{if any
  neighbor is parasitized ($p$)} 
\nonumber \\
r_{h \to e} & = & 1 \\
r_{e \to p} & = & 0 \nonumber \\
r_{p \to h} & = & \varepsilon \nonumber \\
r_{p \to e} & = & \alpha \nonumber
\label{HPmodel}
\end{eqnarray}
As in the {\em SIS} model defined above the decay of the host or first
kind infection sets the time scale ($r_{h \to e} \! = \! 1$). In
biological systems $\alpha > 1$ (even $\gg 1$) if the parasite  
affects the health of the host. A benefit would mean $\alpha < 1$. We
shall consider cases in which the parasite  virtually does not die
``on its own'' but only when the host is killed, i.e.\ the case $0
\approx \varepsilon \ll 1,\lambda, \alpha \mu, \alpha$.

\begin{figure}
\begin{center}
\includegraphics[width=3cm]{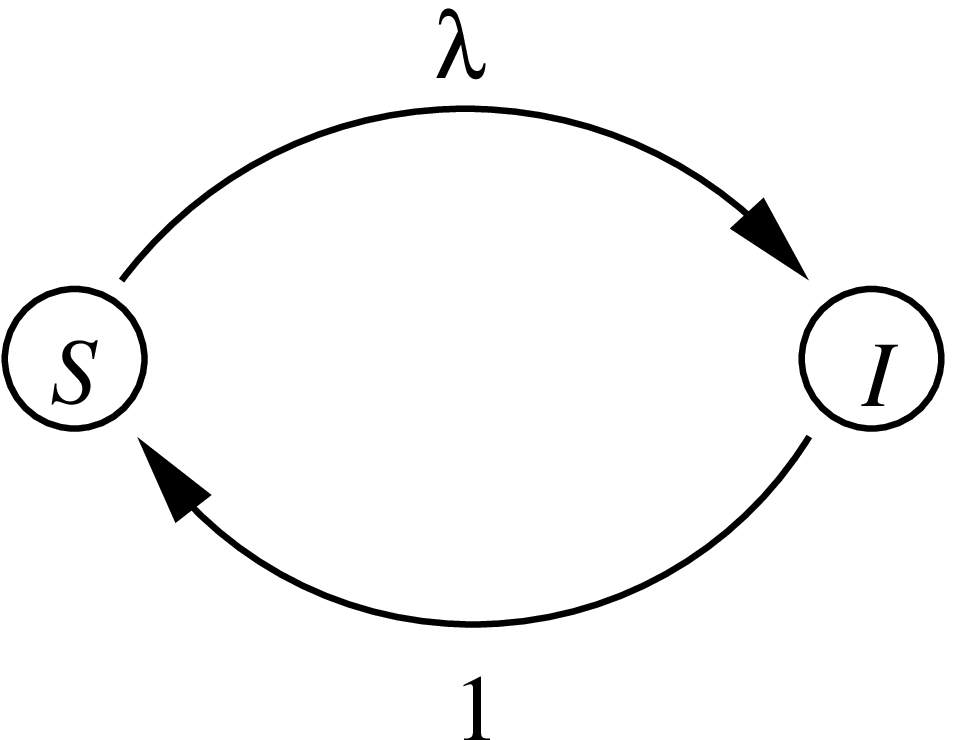}
\hspace{.5cm}
\includegraphics[width=3cm]{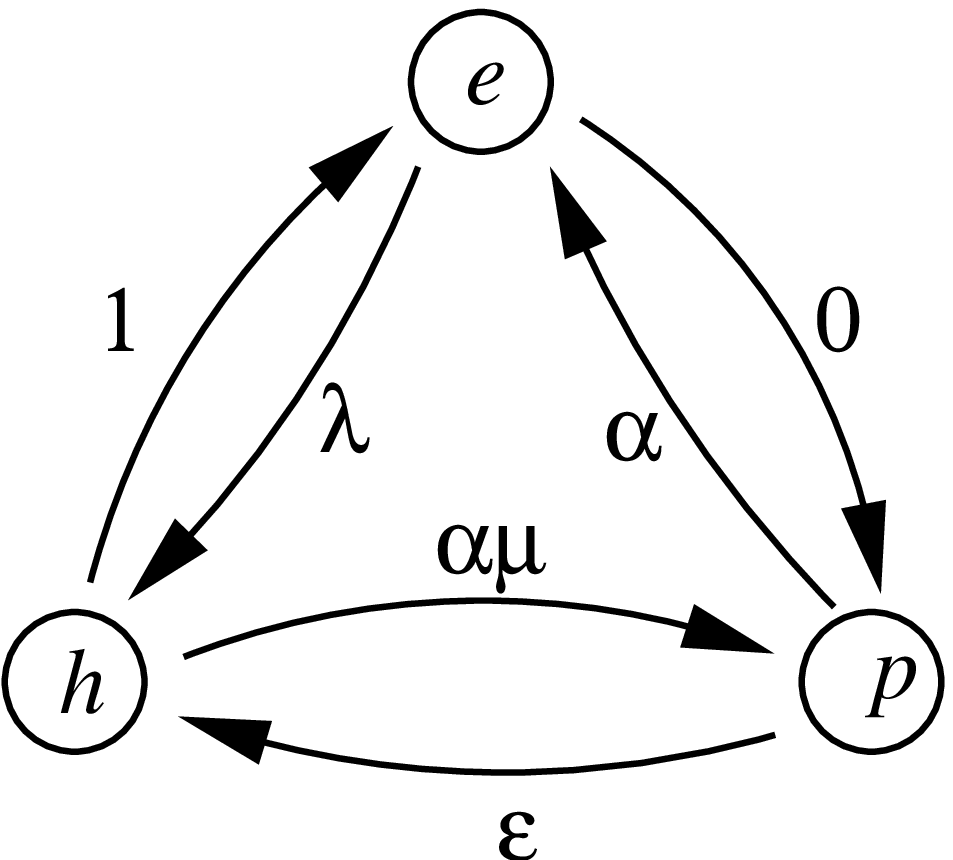}
\end{center}
\label{rates}
\caption{States and rates.}
\end{figure}
In Section \ref{MFD} we present approximate analytical solutions
following \cite{PSV} to the the model of Eqs.~(\ref{HPmodel}) which
are compared to Monte Carlo simulations in Section
\ref{MC}. Particular interest lies in parasite extinction and its
dependence on the parasite spreading rate $\alpha \mu$. But first we
define the types of graphs used in our simulations and calculations.

\subsection{Graphs}
\label{sec:graphs}
We study the population dynamics of the HP--model on two types of
graphs, on Bethe lattices and on scale free Barab\'asi--Albert (BA)
graphs, in their standard version \cite{BA}. A Bethe lattice of
coordination number $z$ is an infinite tree, where each node has $z$
neighbors. When constructing a finite lattice, or Cayley tree,
starting from a central node with $z$ neighbors and adding $z \! - \!
1$ new neighbors to each boundary node, the number of boundary nodes
grows exponentially. It therefore remains a finite fraction of the
total number of nodes in the finite tree, which makes this
construction unsuitable for Monte Carlo simulations.

This difficulty can be overcome by a slight modification \cite{DSS}
where a sparse homogeneous graph that closely approximates the Bethe
lattice without any boundary nodes is constructed. Take $L$ nodes and
label them by integers from 1 to $L$. Connect node $i$ to node $(i+1)$
for each $i$ and connect node 1 to node $L$. Construct $(z-2)$
independent random pairings of the nodes (an easy way to construct
pairings is to sort the nodes randomly and pair the   first node of
this new order with the second one etc.) of the nodes and add an edge
for each pair. By this procedure, we get a graph in which each node is
of degree $z$. For large enough graphs, the loops are negligible
\cite{DSS} and this is a sufficient approximation of a Bethe lattice.

Here, we also use the standard version of Barab\'asi-Albert graphs
(BA-graphs)
\cite{BA}. These are constructed as usual. New nodes are added
one by one 
connecting them with $m \! = \! 3$ links to the previous ones. 
From these, the neighbors are chosen with a probability proportional to
their respective number of links (preferential attachment). 
By this construction highly linked
nodes are likely to obtain even more neighbors as the graph grows,
which results in a ``fat tail'' distribution of probabilities for a
node to have coordination number $k$, $P(k) \sim k^{-3}$ \cite{BA}.
The BA-graphs have very weak degree correlations, i.e. the 
conditional probability for a node of degree $k$ to have a 
neighbor with $k'$ is rather trivial 
\cite{ab02} compared to many other models and real networks.

\section{Mean field and doublet approximation}
\label{MFD}

\subsection{Bethe lattice}

\subsubsection{Singlet (mean field) approach}

In this subsection we extend the known solution for the SIS model on a
Bethe lattice \cite{PSV} to the HP model.  $\rho_h$ and $\rho_p$
denote the density of hosts and parasites, respectively. For
simplicity we consider the limit $\varepsilon = 0$, so parasitized
patches do not supply host individuals to neighboring empty
patches. The rate equations for the densities can be written as
\begin{eqnarray}
\partial_t \rho_h & = & - \rho_h + \lambda \left( 1 \! - \! \rho_p -
\! \rho_h\right) \; \Theta - \alpha \mu \; \rho_h \; \Phi
\nonumber \\ 
\partial_t \rho_p & = & - \alpha \; \rho_p + \alpha \mu \;
\rho_h \; \Phi
\end{eqnarray}
with $\Phi = 1 - (1 \! - \! \rho_p)^z$ and $\Theta = 1 - (1 \! - \! \rho_h)^z$.

In the absence of parasites the host population follows the dynamics
of a SIS model. The trivial state $\rho_h = 0$ is stable for
$\lambda \le 1/z$ and unstable otherwise. In other words, 
the host population can survive only for $\lambda > 1/z$.

Similarly, the pure host phase is stable if parasites cannot invade, 
i.e., if the growth rate of a small parasite population is smaller 
than its death rate,
\begin{equation}
\alpha \; \mu \; z \; \rho_h < \alpha \; \; \; \mbox{or} \; \;
\; \mu < \frac{1}{z \rho_h} \equiv \mu^{\rm crit}.
\end{equation}
From this formula it can be seen that $\mu^{\rm crit} \to
\infty$ as $\rho_h \to 0$, so there is no ``tricritical point'' in the
phase plane, beyond which a direct transition from the absorbed state
to the coexistence state can be seen. The phase diagram is drawn in
Fig.~\ref{fig:bethe-lattice-phasediagram}.

\subsubsection{Doublet approach}
The singlet approach neglects occupancy correlations between adjacent
nodes. The next logical step is a pair or doublet approximation which 
explicitly treats the joint
probabilities to find two unparasitized hosts next to each other
($P_{hh}$), a healthy host next to a parasitized one ($P_{hp}$), and
two parasitized next to each other ($P_{pp}$)
in addition to $\rho_h$ and $\rho_p$.
This approximation is
widely used, we want to emphasize its application to a spatially
uniform insect host--parasitoid model \cite{SMS,OSBH}, to the contact process
in a one--dimensional chain \cite{MD} and in general over a wide
class of models \cite{BK}.

The approximation uses the probabilities $P_{\sigma \sigma'}$ to find
the nodes adjacent to a randomly picked bond in states $\sigma$ and
$\sigma' \in \{ e, h, p \}$, as well as the conditional probabilities
$\rho_{\sigma | \sigma'}$ to find a randomly chosen  
nearest neighbor of a
$\sigma'$--node in state $\sigma$. Three--point and higher
correlations are neglected, so the conditional probabilities to find a
$\sigma$--node next to a $\sigma'$--node which is itself linked to
a third node with state $\sigma''$ are approximated by
\begin{equation} \label{eq:doublet_approximation}
\rho_{\sigma | \sigma' \sigma''} \approx \rho_{\sigma | \sigma'}
\; \; \; \; \; \forall \; \sigma''.
\end{equation}
From there one obtains the rate equations
\begin{eqnarray}
\partial_t \rho_h & = & \Bigl[ -1 - z \alpha \mu \rho_{p|h} + z
  \lambda \rho_{e | h} \Bigr] \; \rho_h \label{eq:rho_h}\\
\partial_t \rho_p & = & \Bigl[- \alpha + z \alpha \mu \rho_{h | p}
  \Bigr] \; \rho_p \label{eq:rho_p} \\
\partial_t P_{hh} & = & - \Bigl[ 2 + 2 (z \! - \! 1) \; \alpha \mu
  \; \rho_{p | h}\Bigr] \; P_{hh} \nonumber \\
 & & + \; \lambda \; \Bigl[ 1 + (z \! - \!
  1) \rho_{h|e} \Bigr] \; P_{he} \label{eq:phh} \\
\partial_t P_{hp} & = & - \Bigl\{ 1 + \alpha + \alpha \mu \;
  \Bigl[ 1 + (z \! - \! 1) \; \rho_{p|h} \Bigr] \Bigr\} \; P_{hp} 
\label{eq:php} \\
 & & + \; (z \! - \! 1) \; \lambda \rho_{h|e} P_{pe} + 2 (z \! - \! 1) \;
  \alpha \mu \rho_{p|h} P_{hh} \nonumber \\
\partial_t P_{pp} & = & -2 \alpha P_{pp} + \alpha \mu \Bigl[1 +
  (z \! - \! 1) \rho_{p|h} \Bigr] \; P_{hp}\\
\partial_t P_{he} & = & - \Bigl\{ 1 + (z \! - \! 1) \alpha \mu
  \rho_{p|h} \nonumber \\ 
 & & + \lambda \Bigl[ 1 + (z \! - \! 1) \; \rho_{h|e} \Bigr]
  \Bigr\} P_{he}\nonumber \\
 & & + 2 (z \! - \! 1) \; \lambda \; \rho_{h|e} P_{ee} + 2 P_{hh} +
  \alpha P_{hp}\\
\partial_t P_{pe} & = & - \Bigl[ \alpha + (z \! - \! 1) \lambda
  \rho_{h|e} \Bigr] \; P_{pe} \label{eq:pep} \\
 & & + \; (z \! - \! 1) \alpha \mu \; \rho_{p|h} P_{he} + P_{hp} + 2
  \alpha P_{pp} \nonumber \\
\partial_t P_{ee} & = & - 2 (z \! - \! 1) \lambda \rho_{h|e} P_{ee} +
  P_{he} + \alpha P_{pe} \label{eq:pee}
\end{eqnarray}
The joint probabilities $P_{\sigma\sigma'}$
can be expressed in terms of the conditional probabilities as
\begin{equation} \label{eq:joint_vs_cond}
P_{\sigma\sigma'} =
\rho_\sigma\rho_{\sigma'|\sigma}(2-\delta_{\sigma,\sigma'}) \, ,
\end{equation}
where $\delta_{\sigma,\sigma'}$ is the Kronecker symbol. 
The factor 2 for $\sigma \neq \sigma'$ reflects the two possible
choices, because $\sigma$ can be on either end of the bond.

There are some subtleties in Eqs.~(\ref{eq:rho_h})--(\ref{eq:pee})
that might not be immediately obvious. In Eq.~(\ref{eq:phh}), for
instance, there is a factor of 2 in the first term. That term
describes a process where an edge connecting two host--sites turns due
to a death of a host into an edge connecting an empty site to a
host--carrying site. The prefactor comes from the fact that this can
happen in two ways, i.e.\ either of the the two hosts can die. For
similar reasons, a prefactor of 2 can also be found in the second term 
of Eq.~(\ref{eq:phh}). However, the rest of the terms in the equation
do not have these prefactors since a similar symmetry does not exist.

In principle Eqs.~(\ref{eq:rho_h})--(\ref{eq:pee}) are solvable in the
steady state. Consider first the SIS model, i.e.~the case without any
parasites. Setting $\mu=0$ and looking at the steady-state of
Eq.~(\ref{eq:rho_h}) immediately yields
\begin{equation} \label{eq:rho_h_intermediate_1}
\rho_{e|h} = \frac{1}{z \lambda} \, .
\end{equation}
Similarly, setting $P_{pe}=0$ in Eq.~(\ref{eq:pee}) and using
the identities $P_{ee}=\rho_e\rho_{e|e}$ and 
$P_{he} = 2\rho_e\rho_{h|e}$ gives
\begin{equation} \label{eq:rho_h_intermediate_2}
\rho_{e|e} = \frac{1}{(z-1)\lambda}
\end{equation}
Expressing $\rho_h$ as
\begin{equation}
\rho_h = \rho_e \frac{\rho_{h|e}}{\rho_{e|h}} = (1-\rho_h)\frac{\rho_{h|e}}{\rho_{e|h}} \, ,
\end{equation}
using the identity $\rho_{e|e} + \rho_{h|e} = 1$,
and plugging in Eqs.~(\ref{eq:rho_h_intermediate_1}) and
(\ref{eq:rho_h_intermediate_2}) finally gives
\begin{equation}
\rho_h = \frac{(z-1)\lambda - 1}{(z-1)\lambda - 1/z} \, , 
\end{equation}
from the numerator of which the critical point follows:
\begin{equation} \label{eq:doublet_crit_lambda}
\lambda_c^{\rm D} = \frac{1}{(z-1)} \, .
\end{equation}
Note that this is different from the mean field result $\lambda_c^{\rm
  MF} = 1/z$. It is also worth noting that rigorous mathematical
results of the contact process \cite{P} give bounds on the critical
point as
\begin{equation}
\frac{1}{z} \le \lambda_c \le \frac{1}{z-1} \, .
\end{equation}

Next consider the boundary between the parasite-absorbing and the 
coexistence phases. Here, hosts live well while parasites are near
extinction. Expanding the steady state solution in the limit of small
parasite population we derive an equation for the phase boundary:
Define two auxiliary quantities $A=\rho_{h|p}$ and $B=\rho_{h|p} +
\rho_{p|p}$. Form an equation for $\partial_t A$ and set it to vanish
since we are looking at the steady-state
\begin{equation} \label{eq:intermediate_A}
\frac{\partial A}{\partial t} \
=\frac{\partial}{\partial t} \frac{P_{hp}}{2\rho_p}
=\frac{1}{2\rho_p} \left( \frac{\partial P_{hp}}{\partial t}-P_{hp}M_p
\right)=0 \,
\end{equation}
where the rate equation (\ref{eq:rho_p})
has been used, and $M_p$ is the Malthusian parameter or
growth rate at low densities of the parasites, i.e.
\begin{equation} \label{eq:malthusian_parasites}
M_p = -\alpha + \alpha\mu z \rho_{h|p} = -\alpha + \alpha \mu z A\, .
\end{equation}
Plugging Eq.~(\ref{eq:php}) into Eq.~(\ref{eq:intermediate_A}),
using Eq.~(\ref{eq:joint_vs_cond}) and the results of 
Eqs.~(\ref{eq:rho_h_intermediate_1}) and
(\ref{eq:rho_h_intermediate_2}) at vanishing parasite population we
arrive at
\begin{eqnarray}
2z(z-1)(1-B)\lambda^2 + & & \nonumber \\
+ \{2 z^2 \alpha\mu A (1-A) + & & \nonumber \\ 
+ 2z[B-1-A(1+2\alpha\mu)]\}\lambda - & & \nonumber \\
- 2(z-1)A\alpha\mu & = & 0 \, . \label{eq:A}
\end{eqnarray}

Similarly, starting at the rate equation for $B$, 
\begin{equation}
\frac{\partial B}{\partial t} = -\frac{\partial \rho_{e|p}}{\partial t}
=\frac{1}{2\rho_p} \left( P_{ep}M_p -\frac{\partial P_{ep}}{\partial t}
\right) \, ,
\end{equation}
using Eqs.~(\ref{eq:pep}) and Eq.~(\ref{eq:joint_vs_cond}) together
with the results of 
Eqs.~(\ref{eq:rho_h_intermediate_1}) and (\ref{eq:rho_h_intermediate_2}), 
one gets
\begin{eqnarray}
2z(z-1)(1-B)\lambda^2 + & & \nonumber \\
+ 2z^2\alpha\mu A(1-B)\lambda + & & \nonumber \\
+ 2z[(B-A)(1-\alpha)-1]\lambda - & & \nonumber \\
- 2(z-1)A\alpha\mu & = & 0 \label{eq:B}
\end{eqnarray}
given that $\lambda \neq 0$.

Now, solve for $A$ in the steady--state version of
Eq.~(\ref{eq:rho_p}), substitute this to Eqs.~(\ref{eq:A}) and
(\ref{eq:B}) and eliminate $B$ from the resulting two equations to get
\begin{equation} \label{eq:doublet_phase_boundary}
\mu = \frac{z(z-1)\lambda^2 + z\alpha\lambda}
{z(z-1)^2\lambda^2 + z[(z-1)(\alpha-1)-\alpha]\lambda 
 + \alpha(1-z)}
\end{equation}
for the phase boundary between parasite-absorbing and co-existence
phases in the $(\lambda,\mu)$-plane. Note that, contrary to the 
mean-field approximation, the phase boundary
defined by this equation does not meet that defined by 
Eq.~(\ref{eq:doublet_crit_lambda}) 
at $\mu\to\infty$
since $\lambda = \lambda_c^D$ is 
not a zero of the denominator of Eq.~(\ref{eq:doublet_phase_boundary}).
It also holds that $\mu_c\to 1/(z-1)$ as $\lambda\to\infty$ so that in
this limit the parasites always find hosts on all nodes and therefore
behave as the SIS model does.

In addition to the solution above we linearize the doublet rate
equations around the previously obtained fixed point with a host 
population and no parasites, i.e.\ $\rho_p = P_{pp} = P_{hp} = P_{ep}
= 0$. Replacing the conditional probabilities $\rho_{\sigma'|\sigma}$
by joint probabilities $P_{\sigma\sigma'}$ as in Eq.~(\ref{eq:joint_vs_cond})
we get a matrix which is of the form
\begin{equation} \label{eq:matrix_coarse}
M = 
\left(
\begin{array}{cc}
M_h & M_{hp} \\ 0 & M_p
\end{array}
\right)
\end{equation}
where $M_h$ governs the stability of the ``host only'' solution,
$M_{hp}$ the effect of a small parasite population on the hosts, and
$M_p$ the growth of parasites at low densities. The block in the lower left
corner is zero since the state without parasites is an absorbing one,
i.e.~a perturbation in the host density cannot reintroduce parasite
population into the system. 

The eigenvalues of a matrix with the structure of
Eq.~(\ref{eq:matrix_coarse}) are just those of $M_h$ and $M_p$,
irrespective of $M_{hp}$. The stability of the host population has
been discussed above, so we are only interested in the (real parts of
the) eigenvalues of the matrix in the following equation,
\begin{equation}
\frac{d}{dt} 
\left(
\begin{array}{c}
\! \rho_p \! \\ \! P_{hp} \! \\ \! P_{pp} \! \\ \! P_{ep} \!
\end{array}
\right) =
\left(
\begin{array}{cccc}
-\alpha & \frac{\alpha z \mu}{2} & 0 & 0 \\
0 & B & 0 & \tilde \lambda \\
0 & \alpha \mu & -2 \alpha & 0 \\
0 & C & 2 \alpha & - \alpha \! - \! \tilde \lambda
\end{array}
\right)
\left(
\begin{array}{c}
\! \rho_p \! \\ \! P_{hp} \! \\ \! P_{pp} \! \\ \! P_{ep} \!
\end{array}
\right)
\label{paraslin}
\end{equation}
with
\begin{eqnarray}
B & = & (z\!-\!1)\alpha \mu (z\lambda \! - \! 1)/(z
\lambda)-1\!-\!\alpha\!-\!\alpha \mu \\
C & = & (z-1)\alpha \mu/(z\lambda) + 1 \; \; \mbox{and} \\
\tilde \lambda & = & (z-1)\lambda-1.
\end{eqnarray}

Note that $\tilde \lambda$ is proportional to the excess over the
critical host growth rate, $\lambda-\lambda_c$.

\begin{figure}
\includegraphics[width=1.0\columnwidth]{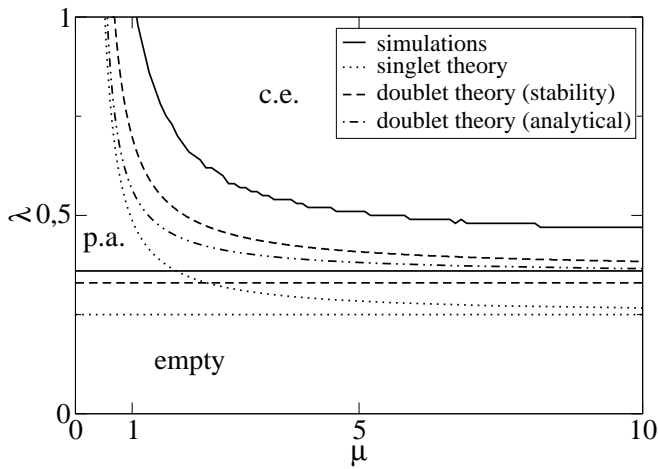}
\caption{Phase diagram for Bethe lattice with $z = 4$ in the
  $(\lambda,\mu)$--plane with $\alpha = 1.2$ and $N=40000$ nodes. 
  Singlet approach (mean
  field) compared to doublet approach (pair approximation) both
  analytically and via stability analysis and Monte Carlo
  simulations. The abbreviations denote the parasite-absorbing (p.a.),
  co-existence (c.e.) and empty phases, the last of which is the phase
  where both populations will eventually become extinct. All three
  solutions are in qualitative agreement with each other. As one
  expects, the pair approximation predicts the need of higher growth
  rates ($\lambda$ and $\mu$) than the singlet approach.}
\label{fig:bethe-lattice-phasediagram}
\end{figure}

The left column of the matrix in Eq.~(\ref{paraslin}) is empty except
for the diagonal element, which gives the first eigenvalue,
$-\alpha$. We therefore restrict ourselves to the remaining $3 \times
3$ matrix. It is straightforward to calculate its eigenvalues
explicitly, from which the phase boundary can be deduced as follows. 
For each fixed $\lambda$, we consider the real part of the largest
eigenvalue of the $3 \times 3$ matrix as a function of $\mu$, and 
find its zero numerically, leading to a point $\mu(\lambda)$ that lies
at the phase boundary. The results are shown in
Fig.~\ref{fig:bethe-lattice-phasediagram} for the case $\alpha = 1.2$
and $z = 4$.

The absence of the tricritical point can be seen easily: As $\lambda
\searrow \lambda_c$ the excess growth rate $\tilde \lambda \searrow
0$, and the matrix becomes lower triangular. All three diagonal
elements yield negative eigenvalues, in particular in this limit $B
\to -\alpha \mu/z - \alpha -1 < 0$. In particular none of the
eigenvalues approaches zero as $\mu\to\infty$, which leads again to
the conclusion that the two phase boundaries do not meet at this
limit.

In comparison to these results the mean field approximation
underestimates the critical values for the spreading parameters. It
does not take into account the clustering of populations, i.e., the
fact that next to a populated site there is likely another one, which
can not be invaded any more. So the possibility for growth is
overestimated.

The phase diagram of the HP model in the $(\mu,\lambda)$-plane obtained 
from both theoretical approaches and from a stochastic simulation using
graph approximation discussed in Sec.~\ref{sec:graphs} is drawn in
Fig.~\ref{fig:bethe-lattice-phasediagram}. In the simulations, rough 
estimates for the phase boundaries were obtained by performing a series of 
simulations with different $\lambda$ for each fixed $\mu$ and observing 
when the population died out. The largest value of $\lambda$ at which the
population dies out is then defined to be the estimate for the position
of the phase boundary. 
From the figure we see that
both analytical solutions are in qualitative agreement with each other
and with the numerical results. A property worth noting of the phase
diagram is the lack of a ``tricritical point'' and thus the phase
boundary between empty and co-existence phases. Consider also that the
singlet approach does reproduce the features of the phase diagram in
the Bethe lattice case.

\subsection{Scalefree graph}

\subsubsection{Singlet approach}

On graphs with non-constant degrees the occupancy of a node depends on its
coordination number. In general, the higher the degree of a node, the greater 
is its tendency to be populated. 
Following Ref.~\cite{PSV} the rate equations for the occupancies
$\rho_h^k$ and $\rho_p^k$ on nodes of degree $k$
can be written as
\begin{eqnarray}
\partial_t \rho_h^k(t) & = & - \rho_h^k(t)  + 
\lambda k \left[ 1-\rho_h^k(t)-\rho_p^k(t) \right]
\Theta_k(\lambda,\mu) \nonumber \\
 & & -\mu \alpha k\rho_h^k(t)\Phi_k(\lambda,\mu) 
\label{eq:rate_scalefree_singlet_hosts}\\
\partial_t \rho_p^k(t) & = & - \alpha \rho_p^k(t) 
+\mu\alpha k\rho_h^k(t)\Phi_k(\lambda, \mu)
\label{eq:rate_scalefree_singlet_parasites}
\end{eqnarray}
where $\Theta_k(\lambda, \mu)$  and $\Phi_k(\lambda, \mu)$ are the
probabilities that a given link points to an infected or a 
parasitized node, respectively. 
In Eq.~(\ref{eq:rate_scalefree_singlet_hosts}) the first term on the RHS
corresponds to the death of the hosts, the second one to the host 
spreading and the third one to parasite spreading, diminishing the
number of sites that carry host but no parasite. 
In Eq.~(\ref{eq:rate_scalefree_singlet_parasites}) the first term on the RHS
describes the death of the parasites while the second one encompasses the 
spreading.
It is known \cite{PSV} that 
there is no epidemic threshold if the distribution of node 
degrees is fat-tailed.

The critical behavior of the HP model as obtained from the mean field
equations above turns out to be incorrect and is in contradiction to the
numerical findings. To see this, consider the rate equations
(\ref{eq:rate_scalefree_singlet_hosts}) and 
(\ref{eq:rate_scalefree_singlet_parasites}) in the limit of small
$\rho$, i.e.\ by a Taylor expansion  in $\rho$. The interaction term
$\mu\alpha\rho_h^k\Phi_k(\lambda, \mu)$ is quadratic in $\rho$
since $\Phi_k(\lambda, \mu) \sim \rho$ and drops out from the
expansion to first order. This, in turn, means that in this limit the
host population behaves as in the SIS model and the parasite
population dies out since its equation only has exponentially decaying
solutions. Furthermore, this rules out the possibility of a zero
epidemic threshold for the parasites, since when the spreading rate
approaches zero also the prevalence does so. This leads to the
aforementioned contradiction. The corresponding numerical results are
presented in Fig.~\ref{fig:pc}.

\subsubsection{Singlet approach with a substantial host population}

Next, we use the singlet approach to look at the behavior of the
parasites when the host population is well established. The
calculation presented here is a straightforward generalization of that
in Ref.~\cite{BPSV}.

The rate equation of the parasites in a Markovian correlated
graph in the singlet approach can be written in the limit of 
small prevalence as 
\begin{equation} \label{eq:rate_parasites}
\frac{\partial}{\partial t} \rho_p^k = -\rho_p^k + 
\alpha\mu \rho_h^k \phi_k \, ,
\end{equation}
where $\phi_k = \sum_{k'}k\Delta_{kk'}\rho_p^{k'}$, 
and $\Delta_{kk'}=P(k'|k)$,
i.e.~the conditional probability that starting from a node of
degree $k$ and following a random edge one is lead to a node of degree
$k'$. For uncorrelated networks, $P(k'|k) = kP(k')/\langle k \rangle$,
where $P(k)$ is the degree distribution of the underlying network.

If the parameters are chosen such that
there are plenty of hosts and that parasites are near extinction 
the feedback coupling of the host population to the parasites can be
neglected and $\rho_h^k$ approximated by a constant vector given
by the solution of the SIS-model. The zero solution $\rho_p^k=0 \,
\forall k$ is always a (formal) solution of the system, so we have to
study its stability. Take Eq.~(\ref{eq:rate_parasites}), 
denote $\rho = (\rho_p^1 \cdots \rho_p^{k_c})^T$ and write the
equation in a matrix form
\begin{equation} \label{eq:rate_matrix}
\frac{\partial}{\partial t}\rho =\left(-I+\alpha\mu
\rho_h^k k \Delta_{kk'}\right)\rho =(-I+\alpha\mu C_{kk'})\rho \, ,
\end{equation}
where $C_{kk'}=\rho_h^k k \Delta_{kk'}$.

Looking at the matrix elements $C_{kk'}$ gives
\begin{equation} \label{eq:generalized_symmetry}
C_{kk'}\frac{P(k)}{\rho_h^k} = C_{k'k}\frac{P(k')}{\rho_h^{k'}} \, 
\end{equation}
where the detailed balance condition of the network \cite{BPS}
\begin{equation} \label{eq:detailedbalance}
kP(k'|k)P(k) = k'P(k|k')P(k')
\end{equation}
has been used. From Eq.~(\ref{eq:generalized_symmetry}) it follows
that $C$ and $C^T$ have the same eigenvalues since, if $v_k$ is any
eigenvector of $C$ corresponding to eigenvalue $\Lambda$, then $v_k
P(k)/\rho(k)$ is an eigenvector of $C^T$ with the same
eigenvalue. This, in turn, has the consequence that the spectrum of
$C$ is real.
Again, the zero solution is unstable, if the matrix $-I+\mu C$
has at least one positive eigenvalue, and the critical value of
$\mu$ is $\mu_{\mathrm{critical}}=\Lambda_M^{-1}$.

Next, use the following corollary \cite{BPSV} of the Frobenius
theorem. Let $A_{kk'}$ be any positive irreducible matrix. Its largest
eigenvalue $\Lambda_M$ can be estimated from below as
\begin{equation} \label{eq:frobenius_corollary}
\Lambda_M \ge \min_k \Bigg\{
\frac{1}{\psi(k)}\sum_{k'}A_{kk'}\psi(k')\Bigg\}
\, ,
\end{equation}
where $\psi(k)$ is an arbitrary positive vector. 
Now, set $\psi(k) = k \rho_h^k$ and $A=C^2$ to get 
\begin{displaymath}
\implies \Lambda_M \ge \min_k\Bigg\{\frac{\sum_{k'} \sum_l 
\rho_h^k k P(l|k) \rho_h^l l P(k'|l) k' \rho_h^{k'}}{k \rho_h^k} \Bigg\}
\end{displaymath}
\begin{equation} \label{eq:lm_estimate}
=\min_k \Bigg\{\sum_l l \rho_h^l P(l|k) \underbrace{\sum_{k'} k' \rho_h^{k'}
P(k'|l)}_{=\overline{k}_{nn}^{(h)}(l,k_c)} \Bigg\} \, .
\end{equation}

Above $\overline{k}_{nn}^{(h)}(l,k_c)$ denotes the average nearest
neighbor degree of such neighbors that carry a host, conditioned that
we are looking at a node of degree $l$. Since the average nearest
neighbor degree of all neighbors $\overline{k}_{nn}(l,k_c) = \sum_{k'}
k' P(k'|l)$ diverges \cite{BPSV2} as $k_c \to \infty$ and $\rho_h^k$
necessarily saturates to a constant value $\rho_h^{k=\infty} \le 1$ with
large $k$, $\overline{k}_{nn}^{(h)}(l,k_c)$ must also diverge at the
same limit. Thus the RHS of Eq.~(\ref{eq:lm_estimate}) diverges,
giving $\Lambda_M \to \infty$ and
\begin{equation} \label{eq:final_result}
\mu_{\mathrm{critical}} \to 0
\end{equation}
at the thermodynamic limit.

\subsubsection{Doublet approach}

Next, we formulate rate equations for a graph with a given degree
distribution and degree-degree--correlations using the doublet
approach or pair approximation. The correlations are included in the
treatment since their use is natural in the context of pair
approximations. The correlated network contains its uncorrelated
counterpart as a special case.

The notation is as follows: $P_{\sigma \sigma'}^{kk'}$ is the
probability that a randomly chosen edge that connects nodes with
connectivities $k$ and $k'$ is such that the state of the node with
connectivity $k$ ($k'$) is $\sigma$ ($\sigma'$), possible states being
$e$, $h$ or $p$. $Q_{\sigma \sigma'}^{kk'}$ is the conditional
probability that a randomly chosen edge that connects nodes with
connectivities $k$ and $k'$ is such that the state of the node with
connectivity $k'$ is $\sigma'$ conditioned that the state of the node
with connectivity $k$ is $\sigma$. 
Let $\Delta_{kk'}$ be as above.

Using the notation above, the rate equations for the SIS model needed for
the present treatment can be written as follows
\begin{eqnarray}
\partial_t \rho_h^k & = & -\rho_h^k + \lambda
\sum_{k'} k \Delta_{kk'} P_{eh}^{kk'} 
\label{eq:rate_host} \\
\partial_t P_{hh}^{kk'} & = & 
-2 P_{hh}^{kk'} + \lambda P_{he}^{kk'} 
\label{eq:rate_host_host}\\
 & & + \lambda \sum_{k''}\Delta_{k'k''} P_{he}^{k''k'} (k'-1)
Q_{eh}^{k'k} \, \nonumber ,
\end{eqnarray}
where in Eq.~(\ref{eq:rate_host_host}) the first term on the right
hand side denotes the process where an infected node gets cured, the
second the process where a node of degree $k$ infects a node of degree
$k'$ and the third the process where a node of degree $k''$ infects a
node of degree $k'$, which in turn has another neighbor of degree $k$
that is infected, turning the edge between the latter two into an edge
connecting two infected nodes.

For the HP model, only one rate equation is needed
for the present treatment, namely that of the
parasite prevalence
\begin{equation} \label{eq:rate_parasite}
\partial_t \rho_p^k = -\alpha\rho_p^k + \mu\alpha
\sum_{k'} k \Delta_{kk'} P_{hp}^{kk'} \, .
\end{equation}

Now consider the steady-state in the SIS model. Multiply
Eq.~(\ref{eq:rate_host}) by $P(k)$ and sum over all $k$ to get
\begin{equation} \label{eq:steady_intermediate1}
\rho_h = \lambda P_{eh} \, 
\end{equation}
$P_{eh}$ is the
fraction of all edges in the network that connect an empty node to one
with host and $\rho_h$ is the average host prevalence in the whole network.

The last term on the right hand side of Eq.~(\ref{eq:rate_host_host})
is positive. Thus in the steady state we can write, leaving out the said
term, 
\begin{equation}
P_{hh}^{kk'} \ge \frac{\lambda}{2} P_{eh}^{kk'} \, .
\end{equation}
Multiplying this by $k P(k) \Delta_{kk'}$, summing over all $k$ and $k'$
and combining with Eq.~(\ref{eq:steady_intermediate1}) we get
\begin{displaymath}
P_{hh} \ge \frac{1}{2}\rho_h \, , 
\end{displaymath}
which implies for the relative density of host--host nearest-neighbor 
pairs that
\begin{equation}
\frac{P_{hh}}{(\rho_h)^2} \ge \frac{1}{2}\cdot\frac{1}
{\rho_h} \to\infty \;\;\; \mathrm{as} \;\;\; 
\rho_h \to 0
\, .
\end{equation}
That is, in the limit of small population, the relative density of 
host--host pairs is enormous. Thus the prevalence correlations in 
nearest-neighbor nodes are also huge. Since the singlet approach
neglects these correlations, this gives reasons to expect that
it is not able to capture the properties in the HP model correctly,
even though it is known that in SIS model it does \cite{BPSV}.

Consider Eq.~(\ref{eq:rate_parasite}) in the steady state. Multiplying
by $P(k)$ and summing over all $k$ gives
\begin{equation} \label{eq:parasite_result}
\rho_p= \mu P_{hp} \, 
\end{equation}
which in turn gives
\begin{equation}
\frac{P_{hp}}{\rho_h\rho_p} = \frac{1}
{\mu\rho_h} \to \infty \;\;\; \mathrm{as} \;\;\;
\rho_p \to 0 \, ,
\end{equation}
since $\mu \to 0$ as $\rho_p\to 0$. 

Eq.~(\ref{eq:parasite_result}) tells that the number of edges through
which the parasite population can spread is proportional to the
parasite prevalence (instead of the product of parasite and host
prevalences). This, in turn, tells that the dynamics of the parasites
is similar to the dynamics of the hosts in the SIS model 
(since in the SIS model the number of edges that can spread the 
population is proportional to the population density in the
steady-state) and serves as
an explanation to the zero threshold of the parasites. 

\section{Monte Carlo Simulations}
\label{MC}

For a numerical comparison we have simulated the host-parasite-model
in Barab\'asi-Albert networks of sizes $L=2^{13}\ldots 2^{21}$ under
the conditions in which $\rho_h \approx 0.30$ and $\rho_p \ll \rho_h$,
i.e.\ with a stable hosts population and parasites close to extinction. 
The simulations are always started with random initial conditions by
giving $25\%$ of the nodes the status $host$ and $5\%$ of the nodes
the status $parasitized$ independently. Then the simulation is run for
a given saturation period of $1000$ MC-steps during which even the
largest system reaches a stationary state. 
Quantities of interest are
then averaged over another $1000$ MC-steps, where one MC-step refers
to the simultaneous (parallel) update event of the state variables of
the nodes. 
The used transition probabilities
$p_{\sigma\sigma'}$ from state $\sigma$ to state $\sigma'$ 
in a single time step are
$p_{eh}=0.012$, $p_{he}=0.05$, $p_{pe}=0.25$ and $p_{hp}$ is varied
in the range from 0.02 to 0.2 to produce the variation in 
$\mu = p_{hp}/p_{pe}$.
This procedure was repeated $N$ times for different
realizations of the graphs with $N$ varying from $N=50$ for $L=2^{13}$
to $N=5$ for $L=2^{21}$.

\begin{figure}
\includegraphics[width=8cm]{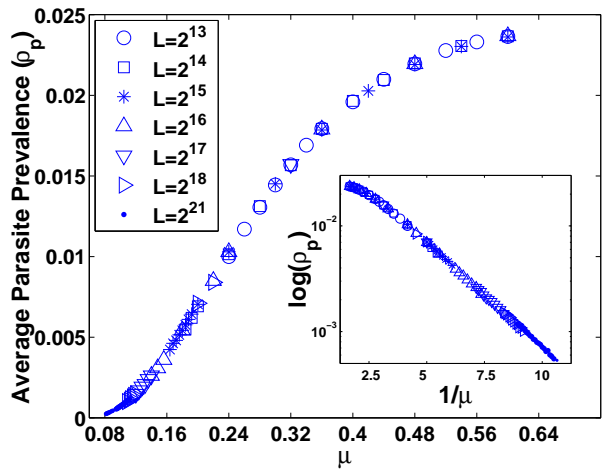}
\hspace{.5cm}
\caption{Average parasite prevalence as a function of its spreading
  parameter $\mu$. The inset corroborates an Arrhenius relation
  $\rho_p \sim \exp(-const/\mu)$, as in the SIS model \cite{PSV}.
  The error bars are smaller than the symbol size.}
\label{fig:rho_vs_p}
\end{figure}

Fig.~\ref{fig:rho_vs_p} shows how $\rho_p$ decays as a function of a
host's parazitation probability parameter $\mu$. Below a size
dependent critical value $\mu_c(L)$ the parasites become extinct
resulting in a left-alone host population obeying dynamics defined by
the SIS-model. For instance when $L=2^{13}$ one may estimate that
$\mu_c \approx 0.26$. The inset in the Figure \ref{fig:rho_vs_p}
strongly suggests that the relationship $\rho_p \sim \exp(-const/\mu)$
is established as in the SIS model \cite{PSV}. 

\begin{figure}
\begin{center}
\includegraphics[width=8cm]{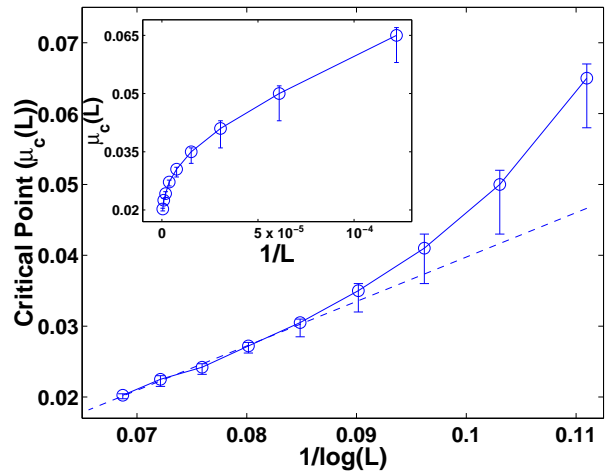}
\hspace{.5cm}
\end{center}
\caption{Scaling of the critical point vs.\ system size. The dashed 
  line works as a guide to the eye and suggest $\mu_c(L) \sim 1/\log(L)$
  as for the SIS model \cite{PSV}.}
\label{fig:pc}
\end{figure}

To track $\mu_c(L)$ more accurately we have studied the extinction
probability ${ \cal P }_{ext}(\mu,L)$ of the parasites during $2000$
MC-steps from different realizations of BA-graphs. The critical point
is then determined to be the highest value of $\mu$ below which the
population dies away in a typical realization of a BA-graph, and the
sizes of the error bars
in $\mu_c$ are estimated from the width of the window in
which ${ \cal P }_{ext}(\mu,L)$ decays from $1$ to $0$.
Fig.~\ref{fig:pc} shows a scaling $\mu_c(L) \sim 1/\log(L)$ in the
region $2^{21}\geq L \geq 2^{16}$, which again compares to the finite
size scaling of the critical threshold in the SIS model \cite{PV}.

\begin{figure}
\begin{center}
\includegraphics[width=8cm]{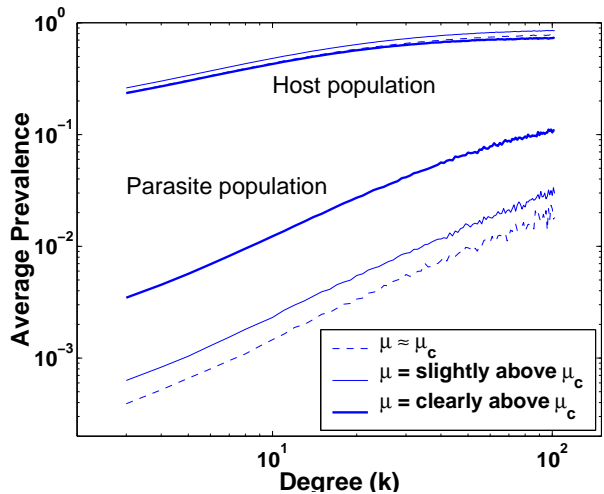}
\hspace{.5cm}
\end{center}
\caption{Average parasite prevalence and its dependence on the nodes
  degree. Here $L=2^{18}$ and only the $\rho_k$ of degree up to
  $k=100$ are shown since the statistics for larger $k$ become worse.}
\label{fig:rhok}
\end{figure}

\begin{figure}
\begin{center}
\includegraphics[width=8cm]{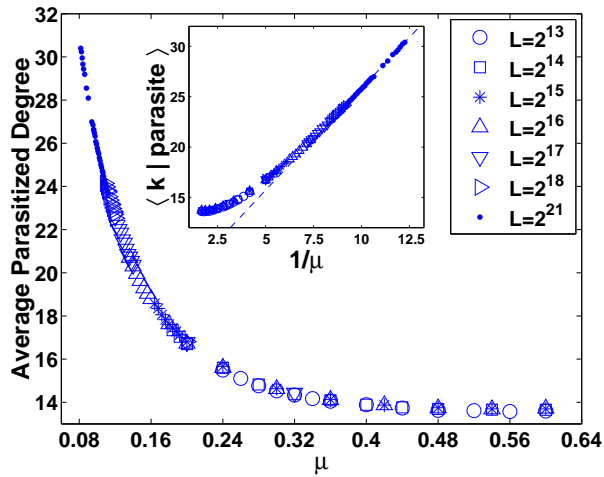}
\hspace{.5cm}
\end{center}
\caption{The expectation value of the degree of a site occupied by the parasites. A
  scaling form for the average degree of parasitized nodes, 
$\langle k | p \rangle$, is found for small $\mu$. The
  straight line in the inset is a guide to the eye.}
\label{fig:Ek}
\end{figure}

Since the probability for a node to become infected depends on its 
degree we next take a look at the parasite prevalence of nodes
of degree $k$ $\rho_p^k$ in
Fig.~\ref{fig:rhok} and the average degree of a site occupied 
by a parasite
$\langle k | p \rangle$ in Fig.~\ref{fig:Ek}.

Fig. ~\ref{fig:rhok} shows that when approaching $\mu_c$ the
relationship $\rho_p^k \sim k$ begins to hold better and better
whereas $\rho_h^k$ does not change remarkably since the host
population is large. In fact, we have noted that, in analogy with the
SIS-model, the scaling of $\rho_p^k$ is not just a matter of
coincidence but reflects the more general presence of the factor
$\rho_p^k \approx 1/(1+\mathrm{const}/k)$ which is proportional to $k$
at small $\mu$, or for large values of the constant. Generally, this
behavior implies that the largest connected component of hosts serves
as a ``scalefree'' graph for the parasites thus partly explaining the
absence of a critical point in the thermodynamic limit.

As $\mu\rightarrow \mu_c$, survival of the parasite population becomes
more and more difficult. Fig.~\ref{fig:Ek} shows a consequence of
this: the parasites do not prefer living in nodes of small degree
anymore but, instead, the average degree of the nodes inhabited by
them increases. In fact, as the inset of Fig.~\ref{fig:Ek} shows, the
scaling $\langle k | p \rangle \sim 1/\mu$ is established. A similar
result should actually also hold for the SIS-model, and is predicted
even by the mean-field equations (\ref{eq:rate_scalefree_singlet_hosts})
and (\ref{eq:rate_scalefree_singlet_parasites}). This in turn follows
from the fact that for a decreasing $\alpha$ the parasite density
begins to saturate only at a higher and higher $k$ (recall that it is
linear in $k$ for small degrees). We have also considered the
autocorrelations of the time-series of the parasite prevalence, in
analogy to Ref.~\cite{epjb}. This decays exponentially with a
time-scale constant that increases as the (pseudo--)critical point
(for $L$ fixed) is approached from above. 

\section{Discussion}

In this paper we have studied a two-population model (``hosts'' and
``parasites''). First, as a preliminary, this problem was considered
on the Bethe lattice. It turns out that the mean-field treatment can
be augmented with the pair approximation. In particular we have been
able to establish the generic form of phase diagram depicted in
Fig.~\ref{fig:bethe-lattice-phasediagram}. This includes no
tricritical point.

The main finding of this work concerns the epidemic threshold of the
parasites, in the presence of a non-zero host population, on
scale-free graphs. Analytical arguments based on the neighbor--pair
probabilities reveal that in full analogy with the SIS model itself,
the threshold is zero in the thermodynamic limit. Numerical
simulations on Barab\'asi--Albert model graphs imply that the
finite-size behavior follows, also, the same scaling, and confirm this
picture. These both findings might be surprising at first sight, due
to the possible complications from correlations. Concomitantly,
correlations in the parasite dynamics are expected to follow the same
picture as in the case of the SIS model. 

A striking feature related to correlated activity
is the ``escape'' of the parasite population to vertices with, on
the average, a high degree, which can actually be explained within the
standard picture of (SIS-type) population behavior as the prevalence
is reduced by changing a control parameter. Due to the non-regular
nature of the scale-free graphs we have not seen any indications of
e.g.\ periodic, or chaotic oscillations that arise in many similar
models on regular lattices \cite{HCM,CHM}. Another possible angle
would be to study contact process--like models \cite{MD}, where the
spreading rate out of a graph vertex to a neighbor would depend
on the degree of the out-vertex, for both parasites and hosts. 
The phase diagram of such model would be the same for the Bethe
case, but for scale-free network one would, in analogy for the
contact process itself \cite{PSC}, expect a {\em finite threshold}
instead of the vanishing one for the SIS model. We have confirmed
this, analytically, but obviously numerical studies would be of
interest.

The results have implications, less for Bethe lattices which serve as
an analytically tractable special case, but possibly for dynamical
processes on real scale-free graphs. Examples can be found from
ecology (metapopulation dynamics), where similar multi-species
scenarios have already been studied. Parasitoids do play a crucial
role for the population dynamics of the endangered butterfly species
{\it melitaea cinxia} in its fragmented habitat on the \AA{}land
islands in the Baltic Sea, which fit less well to single species
models \cite{NH,LH}. Due to the distribution of patch sizes and
distances between them, the corresponding network model has a large
tail degree distribution \cite{epjb}. Whether a given patch is
populated by hosts only or also by parasitoids depends on its local
connectedness. At least qualitatively the observations agree with
those in Fig.~\ref{fig:Ek} and more systematic studies can be
envisioned. The spreading rate depends on distance between and sizes
of patches, in a nontrivial way \cite{HAM}. If one translates the
underlying landscape to a network model, the resulting spreading rates
may depend strongly or weakly on the degree of the emitting node,
i.e.\ lie somewhere in the range between a generalized contact process
and SIS--type models. Thus the limit considered by \cite{PSC} may well
be relevant in certain ecological systems.

Another field of examples is epidemiology and vaccination
strategies. Knowledge on non--trivial network structures in disease
transmission can be used for vaccination (see e.g.\ \cite{OG}) or
outbreak prediction, e.g.\ \cite{MPN}, and also the importance of
superinfections has been documented (see e.g.\ a seminal work in
an evolutionary context \cite{NM}). Our ansatz is an attempt to
combine both points of view. From the scale-free network viewpoint
the fundamental idea of concentrating the effort on nodes with a 
high $k$ is valid here as well \cite{psv2,dalb}; 
consider in particular the ``escape''
of parasites close to extinction mentioned above. To fight parasites
one needs, as well, to avoid random immunization.
In this context another paradigmatic
model is the ``Susceptible-Infected-Removed'' (SIR) model which
is a variant of ordinary percolation. By taking in the HP
model the right combination of limits for the parameters (essentially,
disallowing recovery to the empty state from the H and P states), one
obtains a variant of the SIR which resembles in such language
``bootstrap percolation'' since the ``R'' (``P'') sites are created
only via contact with a neighbor in ``R''. One should thus take note
of possible generalizations of the HP model using similar recipes as
can be applied to the SIR-style ones \cite{DW}.

In the case of the SIS model, the cross-over (or the time-dependent
picture) to the steady-state turns out to be interesting, which might
be worth looking at here as well \cite{BBPSV}. Another practical case
related to this might be, say, viruses spreading as attachments to
emails on the Internet \cite{NFB}, where again one is confronted with
a dynamical graph (of email connections) on top of a larger one
(internet). Finally, we would like to point out that our work could be
extended to other similar multi-species models. An example would be a
hierarchy of contact processes ($A \rightarrow B$, $B\rightarrow C
\dots$) \cite{THH,GHHT}.
\\

{\bf Acknowledgments.} We thank S.~Zapperi and J.~Lohi for stimulating
discussions. An anonymous referee has brought our attention to a small
but important mistake in our calculations. This work was supported by
the Academy of Finland through the Centre of Excellence program (M.A.,
M.P., V.V.) and Deutsche Forschungsgemeinschaft via SFB 611
(M.R.). M.R.\ thanks Helsinki University of Technology for kind
hospitality.

\end{document}